\documentclass{PoS}

\title{Thermodynamics in $2+1$ flavor QCD with improved Wilson quarks by the fixed scale approach}

\ShortTitle{Thermodynamics by the fixed scale approach}

\author{\speaker{T. Umeda}\\
        Graduate School of Education, Hiroshima University, Hiroshima 739-8524, Japan\\
E-mail: \email{tumeda@hiroshima-u.ac.jp}}

\author{S. Aoki\\
        Faculty of Pure and Applied Sciences, University of Tsukuba, Tsukuba, Ibaraki 305-8571, Japan\\
Center for Computational Physics, University of Tsukuba, Tsukuba, Ibaraki 305-8577, Japan}

\author{S. Ejiri\\
        Graduate School of Science and Technology, Niigata University, Niigata 950-2181, Japan}

\author{T. Hatsuda \\
        Theoretical Research Division, Nishina Center, RIKEN, Wako 351-0198, Japan}

\author{K. Kanaya\\
        Faculty of Pure and Applied Sciences, University of Tsukuba, Tsukuba, Ibaraki 305-8571, Japan}

\author{Y. Maezawa\\
        Physics Department, Brookhaven National Laboratory, Upton, New York 11973, USA}

\author{H. Ohno \\
        Fakult\"at f\"ur Physik, Universit\"at Bielefeld, D-33615 Bielefeld, Germany}

\author{(WHOT-QCD Collaboration)}

\abstract{
We study thermodynamic properties of 2+1 flavor QCD with improved Wilson quarks coupled with the RG improved Iwasaki glue, using the fixed scale approach.
We present the results for the equation of state, renormalized Polyakov loop, and chiral condensate.
}

\FullConference{The 30 International Symposium on Lattice Field Theory - Lattice 2012,\\
		June 24-29, 2012\\
		Cairns, Australia}

\begin{document}

\section{Introduction}
Recent progress of computer algorithms and high performance computers
enabled us to perform realistic simulations of $2+1$ flavor QCD with the physical
quark masses \cite{Aoki:2009ix}.
Since the equation of state (EOS) is one of the most expensive calculations 
in lattice QCD, 
realistic simulations for the EOS
has been performed only with staggered-type quarks 
\cite{Cheng:2009zi,Borsanyi:2010cj,Borsanyi:2012uq,Bazavov:2012bp}, 
whose theoretical basis such as 
locality and universality are, however, not well established. 
Therefore, to check the validity of these results it is important to compare 
with those obtained using theoretically sound lattice quarks, 
such as the Wilson-type quarks.
To this end we are pushing forward a project to study the thermodynamics of 2+1 flavor QCD, using a 
nonperturbatively improved Wilson quark action coupled to a RG-improved 
Iwasaki gauge action \cite{PTEP}.

To reduce the high computational cost with Wilson-type quarks, we have proposed the fixed-scale approach, in which we vary temperature 
$T$ by changing temporal lattice size $N_t$ at a fixed lattice spacing $a$ \cite{Umeda:2008bd,Umeda:2012er,PTEP}.
Because we fix the coupling parameters for all $T$'s, one zero-temperature simulation at this point can be used in common to renormalize observables at all $T$'s.
Together with other good features of the fixed-scale approach, we can reduce the computational cost for zero-temperature simulations, which is a big burden in the more conventional fixed-$N_t$ approach.
In the fixed-scale approach, the trace anomaly $\epsilon -3p$ is calculated as usual at each temperature.
To calculate the pressure non-perturbatively, we have developed the `$T$-integration method' \cite{Umeda:2008bd}:
\begin{eqnarray}
\frac{p}{T^4} = \int^{T}_{T_0} dT \, \frac{\epsilon - 3p}{T^5}
\label{eq:Tintegral}
\end{eqnarray}
with $p(T_0) \approx 0$.

In this report we present the results of EOS, Polyakov loop, and chiral condensate 
in $2+1$ flavor QCD using the fixed-scale approach. 
Calculation of the beta-functions is updated from the previous reports \cite{Kanaya2009,Umeda:2010ye}.
Chiral condensate with Wilson-type quarks is known to have a severe divergence due to the explicit chiral symmetry breaking of the Wilson term \cite{Bochicchio:1985xa,Giusti:1998wy}.
We renormalize the Polyakov loop and chiral condensate taking the advantage of the fixed-scale approach that the renormalization is common to all temperatures.

The lattice set-up is given in the next section.
Details of beta-functions and results of EOS are presented in Sec.3 and 4.
The Polyakov loop and the chiral condensate are discussed in Sec.5 and 6.
The report is summarized in the last section.
Part of the results including the details of the EOS calculation was published recently in \cite{Umeda:2012er}.

\section{Lattice setup}
\label{sec:setup}
A good feature of the fixed-scale approach is that we can use high-precision zero-temperature configurations on fine lattices, openly available on ILDG etc., for the renormalization of observables at all temperatures.
As the zero-temperature configurations, we choose those of a 2+1 flavor QCD spectrum study with 
improved Wilson quarks by the CP-PACS+JLQCD Collaboration \cite{Ishikawa:2007nn} in the study.
The action $S=S_g+S_q$ is a combination of the
RG-improved gauge action $S_g$ and the clover-improved Wilson quark action
$S_q$,
\begin{eqnarray}
S_g &=& -\beta\left\{ 
\sum_{x,\mu>\nu}c_0W^{1\times 1}_{\mu\nu}(x)
+\sum_{x,\mu,\nu}c_1W^{1\times 2}_{\mu\nu}(x)
\right\},\\
S_q &=& \sum_{f=u,d,s}\sum_{x,y} \bar{q}_x^f D_{x,y}q_y^f, \\
D_{x,y} &=& \delta_{x,y}-\kappa_f
\sum_\mu\{ (1-\gamma_\mu)U_{x,\mu}\delta_{x+\hat{\mu},y}
+(1+\gamma_\mu)U^\dagger_{x-\hat{\mu},\mu}\delta_{x-\hat{\mu},y}
\}
-\delta_{x,y}c_{SW}\kappa_f\sum_{\mu>\nu}\sigma_{\mu\nu}
F_{\mu\nu},
\nonumber
\end{eqnarray}
where $c_{SW}$ is non-perturbatively determined as a function of $\beta$ \cite{Aoki:2005et}.
Among their simulation points, 
we choose $\beta=2.05$, $\kappa_{ud}=0.1356$ and 
$\kappa_s=0.1351$, which is the smallest lattice spacing
[$a\simeq 0.07$ fm with a scale set by $r_0=0.5$ fm] and the lightest $u$ and $d$ quark masses  
[$m_\pi/m_\rho=0.6337(38)$].
The $s$ quark mass corresponds to $m_{\eta_{ss}}/m_{\phi}=0.7377(28)$.
The lattice size is $28^3 \times 56$.

Using the same coupling parameters as the zero-temperature simulation, 
we have generated finite temperature configurations on 
$32^3\times N_t$ lattices with $N_t=16$, 14, $\cdots$, 4 \cite{Umeda:2012er}.
Our range of $N_t$ corresponds to the range $T=174$--696 MeV.
The pseudo-critical point is expected to be $N_t \approx 14$. 


\section{Beta functions}
\label{sect:beta}

To evaluate the trace anomaly, we need the information of beta functions $a(d\beta/da)$ and 
$a(d\kappa_f/da)$ ($f=ud$ and $s$). 
In this study, we define the lines of constant physics (LCP's) by $m_\pi/m_\rho$ and $m_{\eta_{ss}}/m_\phi$ at $T=0$.
The beta functions are determined nonperturbatively through the coupling parameter 
dependence of zero-temperature observables on a LCP.
We use the data of $am_\rho$, $m_\pi/m_\rho$ and $m_{\eta_{ss}}/m_\phi$ at 
30 simulation points of the CP-PACS+JLQCD zero-temperature configurations 
\cite{Ishikawa:2007nn} to extract the beta functions.

\begin{figure}[bt]
  \begin{center}
    \begin{tabular}{ccc}
    \includegraphics[width=65mm]{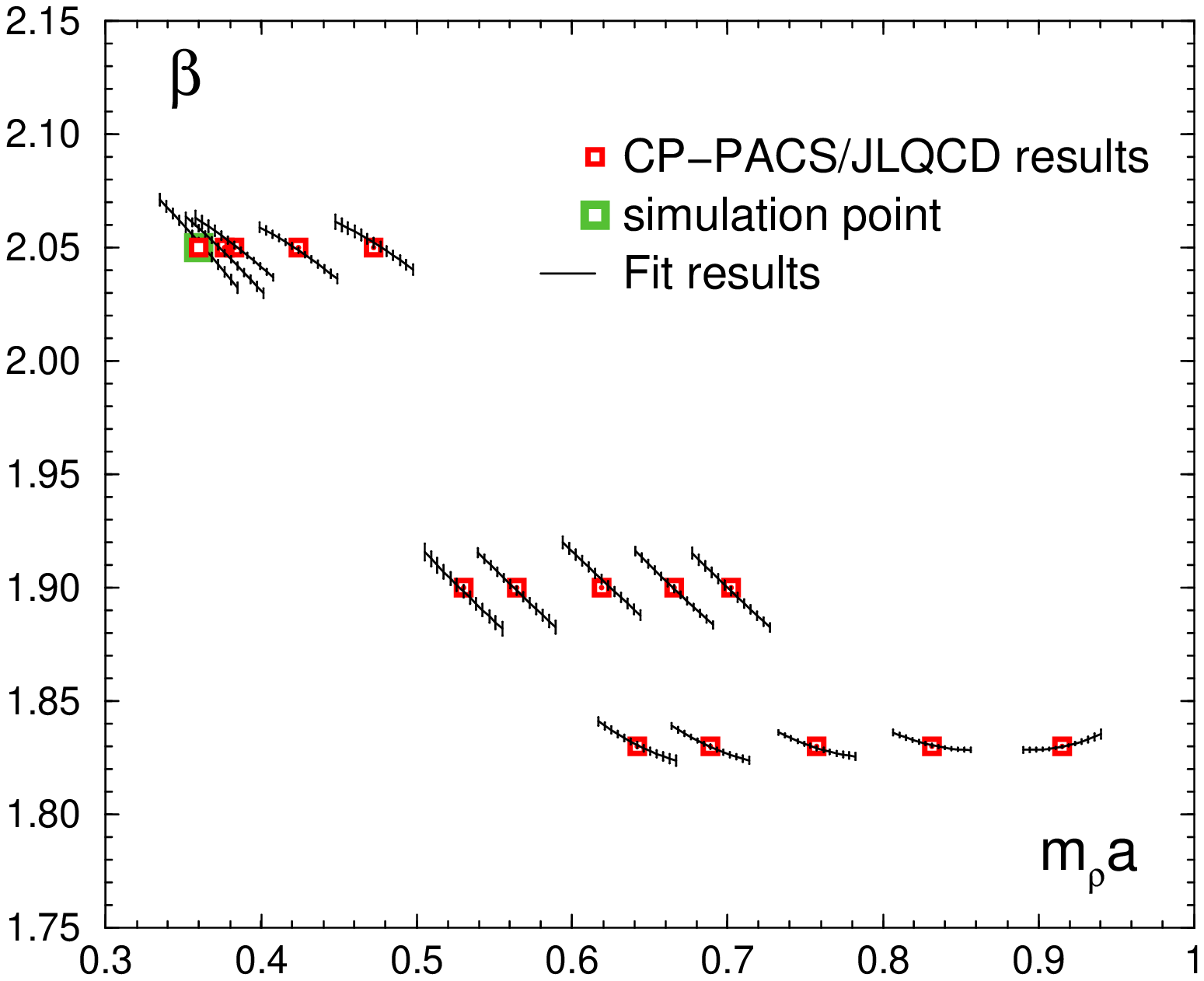}& 
    \includegraphics[width=65mm]{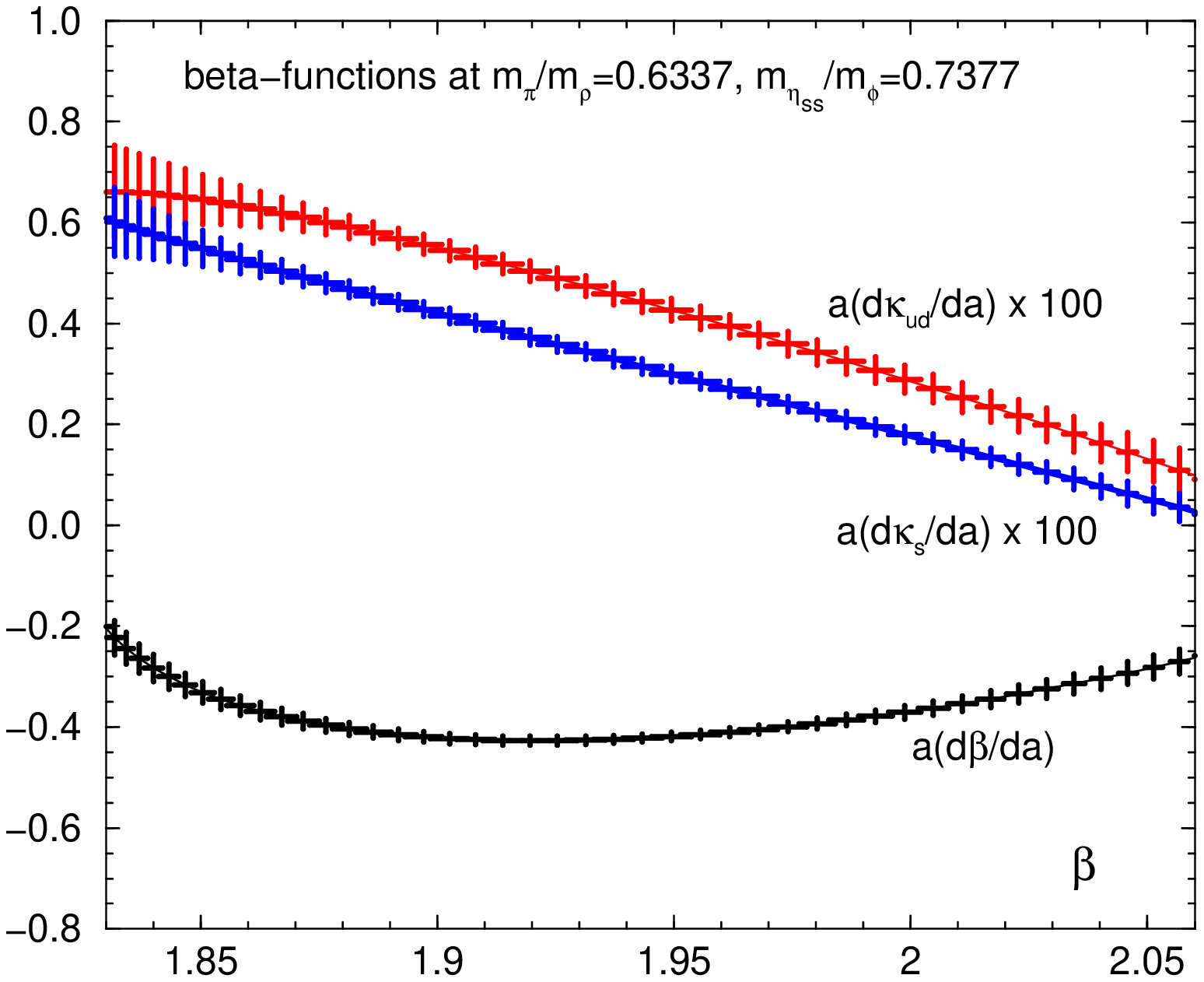} 
    \end{tabular}
\caption{{\em Left}: 
    The global fit for $\beta$ as a function of $m_\rho a$.
    Square symbols show the simulation points by CP-PACS+JLQCD. 
    Solid curves are the fit results around these simulation points with corresponding 
    $m_\rho/m_\pi$ and $m_{\eta_{ss}}/m_\phi$.
    Open green square is the simulation point in this study.
    To avoid a too busy plot, only half of the data points are shown.
    {\em Right}: 
     Beta functions on our LCP ($m_\pi/m_\rho=0.6337$, $m_{\eta_{ss}}/m_{\phi}=0.7377$) 
     as functions of $\beta$. The scale is set by $am_\rho$. 
     Beta functions for $\kappa_{ud}$ and $\kappa_{s}$ are magnified by factor 100.
   }
    \label{fig:bfunc}
  \end{center}
\end{figure}

We have first tried to evaluate the beta functions by `the inverse matrix method',
{\it i.e.}, fit the data of $a m_\rho$, $m_\pi/m_\rho$ and $m_{\eta_{ss}}/m_\phi$ as functions of the coupling parameters $\beta$, $\kappa_{ud}$ and $\kappa_s$, and then invert the matrix of the slopes to obtain the beta functions \cite{Kanaya2009}.
It turned out that a high statistics and many data points are required to suppress the errors in $a (d\kappa_f/da)$ because these beta functions have much smaller magnitude than $a(d\beta/da)$ and thus the error in the latter contaminates those of the formers through the matrix inversion procedure.

To avoid the matrix inversion, we thus adopt `the direct method', in which we fit the coupling parameters, $\beta$, $\kappa_{ud}$ and $\kappa_s$ as functions of the observables $am_\rho$, $m_\pi/m_\rho$ and $m_{\eta_{ss}}/m_\phi$. 
Consulting the overall quality of the fits, we have refined the fit function and the fit range from our previous report \cite{Umeda:2010ye}.
We now fit with the third order polynomial \cite{Umeda:2012er}.

The result of the global fit for $\beta$ around each simulation point is shown in Fig.\ref{fig:bfunc}  
as a function of $am_\rho$.
The fit works well with $\chi^2/{\rm dof}=1.63$, 1.08, and 1.69 for the fit of $\beta$, $\kappa_{ud}$, and $\kappa_s$, respectively, where the degree of freedom is 10 for each coupling parameter.
The resulting beta functions for our LCP are shown 
in Fig.~\ref{fig:bfunc} as functions of $\beta$.
As a variable to set the scale, we may alternatively adopt $am_\pi$, $am_{K}$ or $am_{K^*}$ instead of $am_\rho$.
We find that the results are roughly consistent with each other.
Taking the results from $am_\rho$ as the central value, we obtain 
\begin{equation}
a\frac{d \beta}{d a} = -0.279(24)(^{+40}_{-64})
,\hspace{5mm}
a\frac{d \kappa_{ud}}{d a} = 0.00123(41)(^{+56}_{-68})
,\hspace{5mm}
a\frac{d \kappa_s}{d a} = 0.00046(26)(^{+42}_{-44})
\end{equation}
at our simulation point,
where the first brackets are for statistic errors, and the second brackets are for systematic errors estimated by the variation of the scale setting.

\section{Equation of state}

\begin{figure}[bt]
  \begin{center}
    \begin{tabular}{cc}
    \includegraphics[width=68mm]{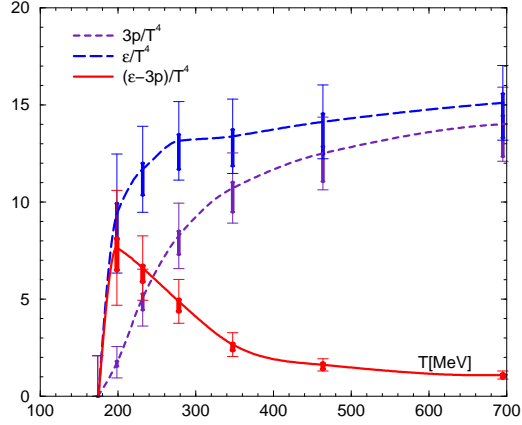}
    \end{tabular}
   \caption{
    Trace anomaly $(\epsilon-3p)/T^4$, energy density $\epsilon/T^4$,
    and pressure $3p/T^4$ in 2+1 flavor QCD. 
    The thin and thick vertical bars represent statistic and systematic errors, 
respectively.
    The curves are drawn by the Akima spline interpolation.
    }
    \label{fig:eos}
  \end{center}
\end{figure}

Using the results of the beta functions, we calculate the trace anomaly $(\epsilon-3p)/T^4$. 
See Ref.\cite{Umeda:2012er} for the details of the calculation. 
In Fig.~\ref{fig:eos}, we show the results of the trace anomaly by a thick red curve. 

Carrying out the $T$-integration of the trace anomaly according to Eq.~(\ref{eq:Tintegral}), 
we obtain the pressure $p$ shown in Fig.~\ref{fig:eos}.
The energy density $\epsilon$ is then calculated from $p$ and $\epsilon-3p$.
Although our errors are still large, our EOS shown in Fig.~\ref{fig:eos} is roughly consistent with 
recent results with highly improved staggered quarks near the physical point 
\cite{Borsanyi:2010cj,Bazavov:2012bp}. 

\section{Renormalized Polyakov loop}
\label{sec:poly}

The Polyakov loop 
$
L = (3N_s^3)^{-1} \sum_{\vec{x}} {\rm Tr}\prod_{x_4} U_{4}(x)
$
measures the quark free energy $F(T)$.
To renormalize $F(T)$, we have to add an additive renormalization constant, which is independent of $T$ and thus common for all $T$'s in the fixed-scale approach.
However, renormalization of $L$ depends on $T$ due to the relation $\langle L \rangle \sim e^{-F/T}$.
We adopt the renormalization scheme of Ref.\cite{Cheng:2007wu}. 
The renormalized Polyakov loop is then given as $L_{\rm ren} = Z_{\rm ten}^{N_t} L$, 
where we set $Z_{\rm ren}=1.4801(90)$ from our potential data obtained at $T=0$ \cite{Maezawa:2011aa}. 

\begin{figure}[tb]
    \includegraphics[width=68mm]{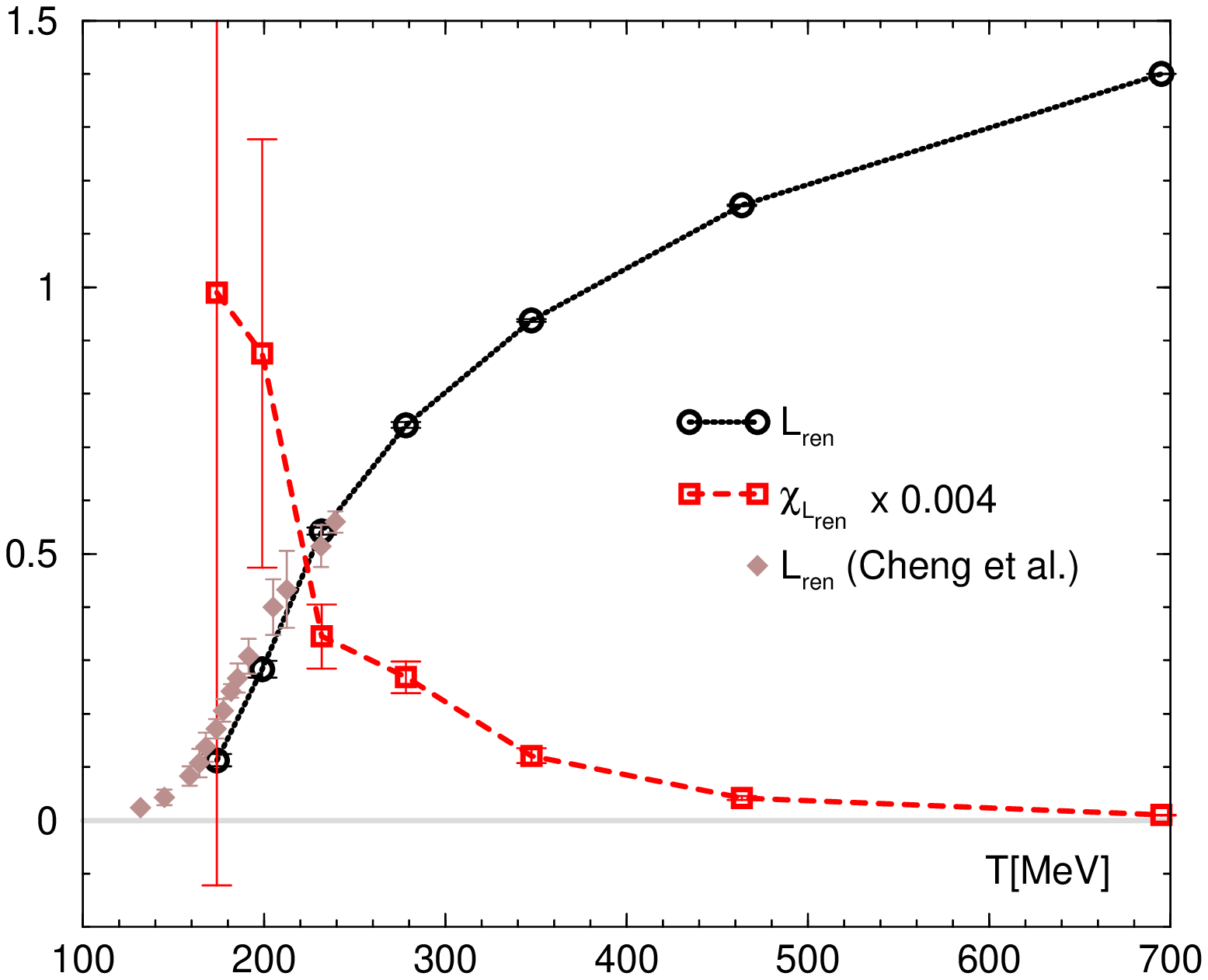}
    \hspace{5mm}
    \includegraphics[width=65mm]{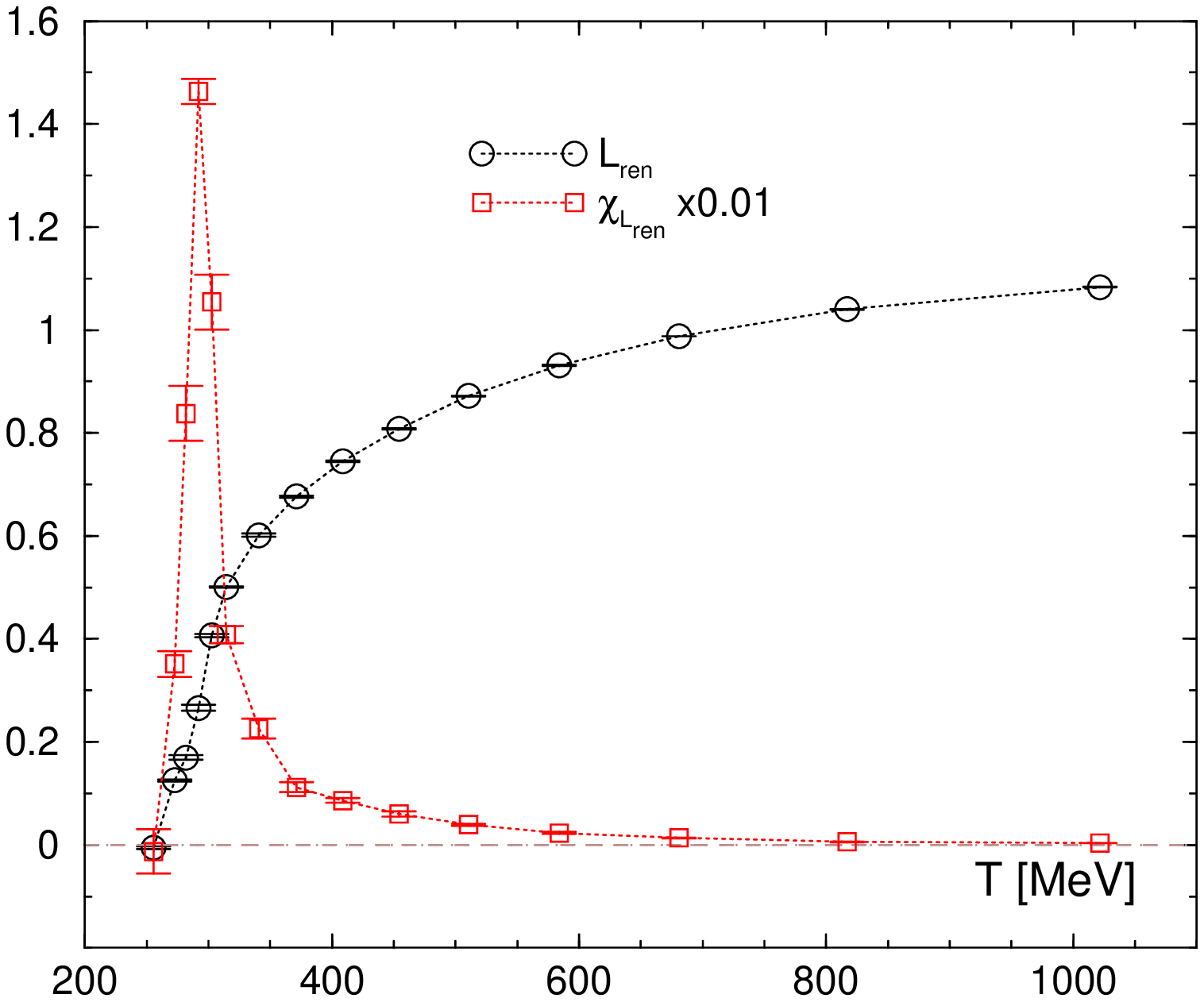}
\caption{Renormalized Polyakov loop and its susceptibility 
as functions of $T$, obtained with the fixed-scale approach.
The left panel shows the results for $2+1$ flavor QCD.
The same quantities obtained in quenched QCD \cite{Umeda:2008kz} are shown in the right panel.
}
\label{fig:poly_sus}
\end{figure}

Our results for $\langle L_{\rm ren} \rangle$ are plotted in the left panel of 
Fig.~\ref{fig:poly_sus}. 
As discussed in \cite{Umeda:2010ye}, the $T$-dependence in these quantities are largely influenced by the renormalization factor.
We note that our $\langle L_{\rm ren} \rangle$ agree well with a result from the p4 staggered quark 
action in the fixed-$N_t$ approach at $N_t = 8$ \cite{Cheng:2007jq}, as shown by shaded symbols in the figure, although our quarks are heavier than theirs.

In the same figure, the results for the susceptibility $\chi_{L_{\rm ren}}$ are also shown. 
We expect the pseudo-critical temperature around 200 MeV.
We do not see a clear peak in $\chi_L$, although a peak at $T \sim 180$--200 MeV is not excluded due to the large errors. 
For comparison, we show, the results for the same quantities obtained in quenched QCD adopting the fixed scale approach in the right panel of Fig.~\ref{fig:poly_sus} \cite{Umeda:2008kz}.
In this case, we confirm a clear peak of the susceptibility at the weakly 1st order deconfinement transition temperature. 
We thus suspect that a peak is hidden in the large errors and/or between the two data points. 
Some reasons of the lower resolution of $T=(N_t a)^{-1}$ in the full QCD study are that $N_t$ is restricted to be even due to the CPS simulation code we used, and that the lattice spacing $a$ is coarser than that in the quenched study.
We also note that the crossover may be weak due to the large quark mass, large $N_t$, small aspect ratio $N_s/N_t$, etc.
We reserve investigation of these points for future study at lower quark masses.

\section{Chiral condensate and its susceptibility}

Finally, let us study the chiral condensate.
Because the chiral symmetry is explicitly broken with Wilson-type quarks,
chiral condensate $\sigma_R$ requires additive and multiplicative
renormalizations \cite{Bochicchio:1985xa,Giusti:1998wy},
\begin{eqnarray}
  \langle \bar{\psi}\psi\rangle (T) = \frac{T}{V}\langle Tr D^{-1} \rangle
  = \frac{\langle\sigma_R(T)\rangle}{Z_{\bar{\psi}\psi}} + c_{\bar{\psi}\psi} .
\end{eqnarray}
Here $c_{\bar{\psi}\psi}$  and $Z_{\bar{\psi}\psi}$ are the renormalization constants.
Because the renormalization is done such that the chiral symmetry is recovered in the continuum limit,
we can alternatively define the chiral condensate in terms of the Ward-Takahashi identity \cite{Bochicchio:1985xa},
\begin{eqnarray}
m_q \, \langle \sum_x P(x)P(0)\rangle (T) 
  = \frac{\langle\sigma_R(T)\rangle}{Z_{PP}},
\end{eqnarray}
where $P(x)$ and $m_q$ are the pseudoscalar density and the quark mass.
Corresponding susceptibilities are given by 
\begin{eqnarray}
\chi_{X} =  
C_{X} \left( \langle \sigma_R^X(T)^2\rangle 
-\langle \sigma_R^X(T)\rangle^2 \right),
~~~ X=\bar{\psi}\psi ~~\mbox{or} ~~\mbox{PP},
\end{eqnarray}
where $C_{\bar{\psi}\psi}=1/Z_{\bar{\psi}\psi}^2$ and $C_{PP} = 1/(m_q Z_{PP})^2$.
Note that the factors $Z_{\bar{\psi}\psi}$, $Z_{PP}$, and $m_q$ are just constants common to all $T$'s in the fixed-scale approach.
Note also that the additive renormalization constant $c_{\bar{\psi}\psi}$ cancels out in the calculation of $\chi_{\bar{\psi}\psi}$.

\begin{figure}[bt]
  \begin{center}
    \begin{tabular}{cc}
    \includegraphics[width=65mm]{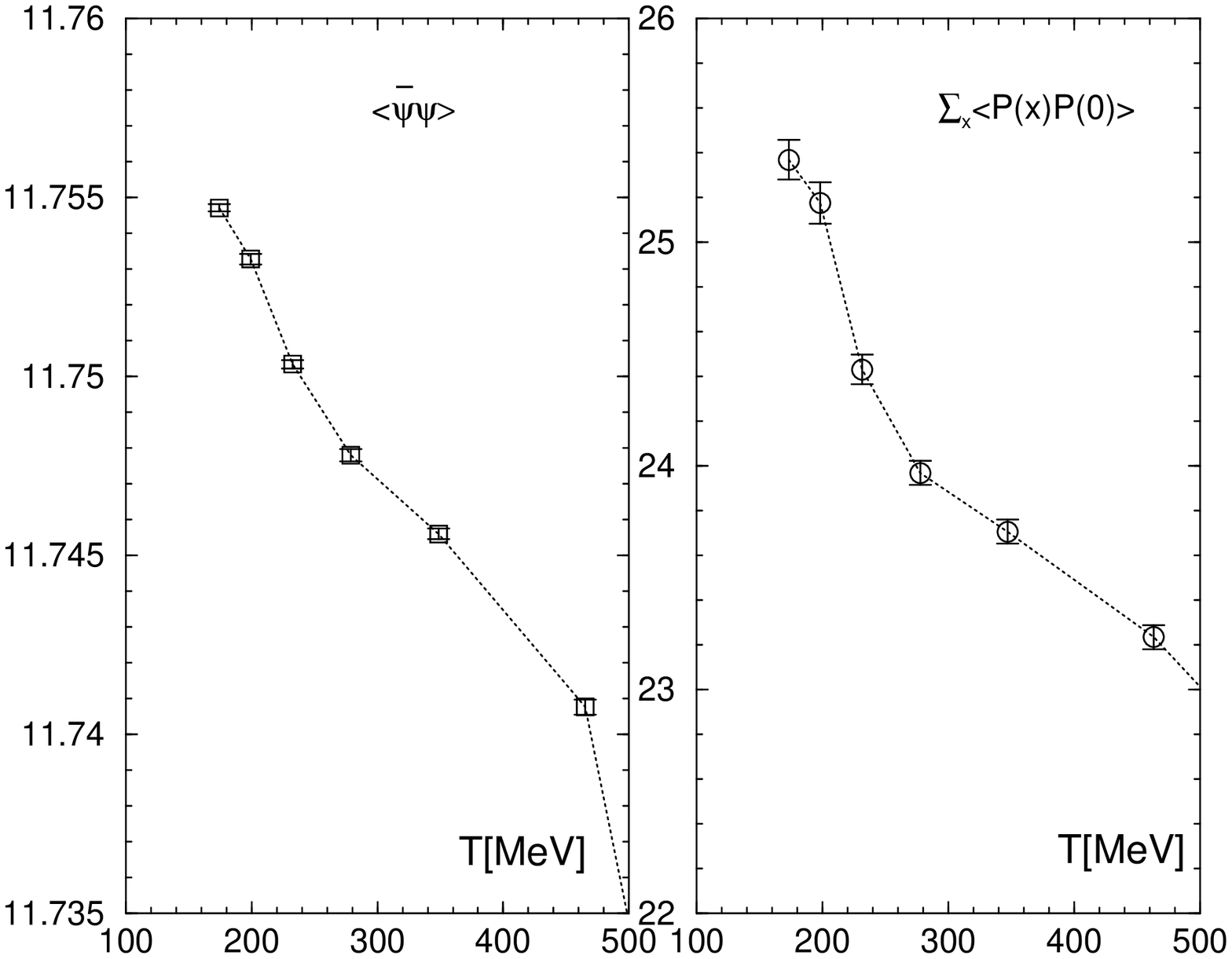}
    \hspace{5mm}
    \includegraphics[width=62mm]{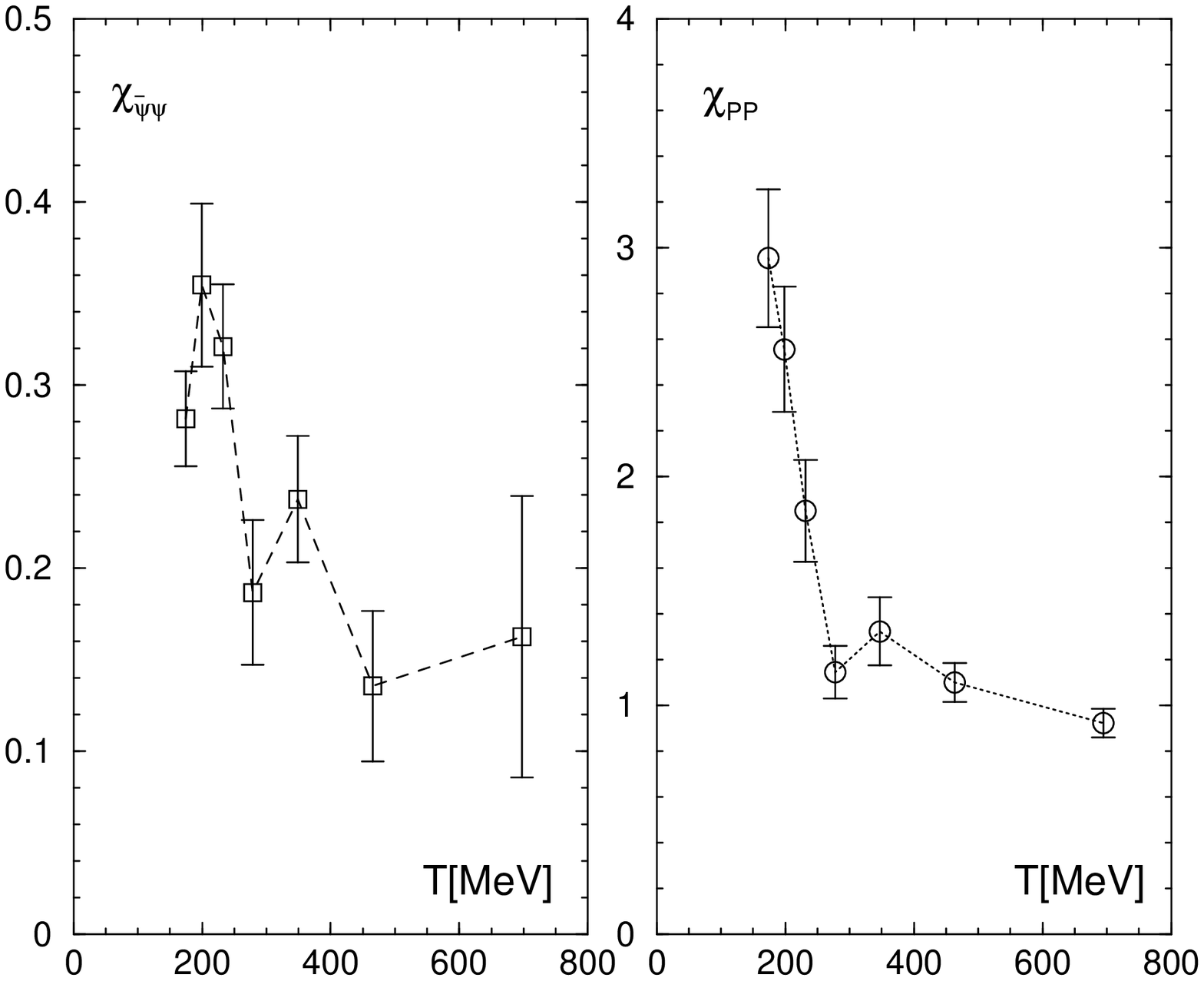}
    \end{tabular}
    \caption{
    Bare chiral condensates (left) and their susceptibilities (right) as a function of temperature.
    }
    \label{fig:chiral}
  \end{center}
\end{figure}

In Fig.\ref{fig:chiral}, the bare chiral condensates and their susceptibilities are plotted.
Ignoring the difference in the overall scale, the chiral condensates and the susceptibilities are consistent with each other within the statistical errors. 
However, a clear peak could not be seen in the susceptibilities, as in the case of the Polyakov loop susceptibility, except for a small peak in the $\chi_{\bar{\psi}\psi}$.

\section{Summary}

Adopting the fixed-scale approach, 
we have calculated the EOS in $2+1$ flavor QCD with non-perturbatively improved Wilson-type quark action combined with the RG-improved Iwasaki gauge action.
To our knowledge, this is the first result of the $2+1$ flavor EOS with Wilson-type quarks.
Although the light quark masses are heavier than their physical values yet, 
our EOS looks roughly consistent with recent results with highly improved staggered 
quarks \cite{Borsanyi:2010cj,Bazavov:2012bp}.
We also studied the Polyakov loop, chiral condensate, and their susceptibilities.
With the current precision of the simulation, however, we could not find a clear signal of the pseudo critical point.
To improve the situation, it is important to achieve a higher resolution in $T$, e.g., by performing simulations at odd $N_t$ and/or by combining data at different scales.
We are planning to test these ideas in a study at the physical point, 
using the on-the-physical-point configurations generated by the PACS-CS 
Collaboration \cite{Aoki:2009ix}.

\vspace{5mm}
We thank the members of the CP-PACS and JLQCD Collaborations for 
providing us with their high-statistics 2+1 flavor QCD configurations.
This work is in part supported 
by Grants-in-Aid of the Japanese Ministry
of Education, Culture, Sports, Science and Technology, 
(Nos.22740168, 
21340049  
23540295  
20340047  
) and the Grant-in-Aid for Scientific Research on Innovative Areas
(Nos.20105001, 20105003 
, 23105706 
).  
This work is in part supported also by the Large Scale Simulation Program of High Energy Accelerator Research Organization (KEK) Nos. 09/10-25, 10-09, (T)11-13 and 12-14.


\end{document}